\documentclass{mnras}
\usepackage{graphicx}

\def\pg{{PG1211+143}}
\def\mkn{{Mkn509}}

\def\xmm{{\it XMM-Newton}}

\def\suzaku{{\it Suzaku}}

\def\et{{et al.\ }}

\newcommand{\ls}{\mathrel{\hbox{\rlap{\hbox{\lower4pt\hbox{$\sim$}}}\hbox{$<$}}}}
\newcommand{\gs}{\mathrel{\hbox{\rlap{\hbox{\lower4pt\hbox{$\sim$}}}\hbox{$>$}}}}

\def\Msun{\hbox{$\rm ~M_{\odot}$}}

\def\H0{{\rm ~km~s^{-1}~Mpc^{-1}}}

\def\et{{et al.}}

\title [Shock cooling]
       {Possible evidence for shock-cooling in the accretion flow of the luminous Seyfert galaxy \pg }
\author[Ken Pounds and Andrew Lobban]
       {Ken Pounds $^{1}$ and Andrew Lobban $^{2}$\\
$^{1}$ Department of Physics and Astronomy, University of Leicester, Leicester, LE1 7RH, UK\\
$^{2}$ European Space Astronomy Centre, Madrid, Spain}

\date{Accepted ; Submitted }
\pagerange{\pageref{firstpage}--\pageref{lastpage}}
\pubyear{2005}
\begin{document}

\label{firstpage}
\pagerange{\pageref{firstpage}--\pageref{lastpage}}

\maketitle

\begin{abstract}
Short-term variability and multiple velocity components in the powerful highly ionized wind of the archetypal UFO \pg\ are indicative of inner
disc instabilities or short-lived accretion events. The recent detection of a high velocity {\it inflow} offered the first direct observational
support for the latter scenario, where matter approaching at a high inclination to the black hole spin plane may result in warping and
tearing of the inner accretion disc, with subsequent inter-ring collisions producing shocks, loss of rotational support and rapid mass infall. Here we identify
a variable continuum component in the same data set, well-modelled by a hot thermal Comptonised spectrum that could represent cooling radiation from the shocked gas.

\end{abstract}


\section{Introduction}
X-ray spectra from an \xmm\ observation of the luminous Seyfert galaxy \pg\ in 2001 identified strongly blue-shifted
absorption lines of highly ionized gas, corresponding to a sub-relativistic outflow velocity of
0.15$\pm$0.01$c$ (Pounds \et\ 2003; Pounds \& Page 2006). Archival data from \xmm\ and \suzaku\ subsequently showed that such ultra-fast,
highly-ionized outflows (UFOs) are relatively common in nearby, luminous AGN (Tombesi \et\ 2010, 2011; Gofford \et\ 2013). While those archival
searches typically reported a single velocity, in the few cases where an AGN was observed repeatedly, the wind velocity was often different.
\mkn\ is the best example from the \xmm\ data archive, with wind velocities of $\sim$0.173c, $\sim$0.139c and $\sim$0.196c, separated by 5 years
and 6 months respectively (Cappi \et\ 2009).

An extended 5-weeks \xmm\ observation of \pg\ in 2014 showed a more complex velocity structure in the highly ionized wind, with three
primary (high column density)  outflow velocities of $v \sim 0.066c$, $v \sim 0.13c$ and $v \sim 0.19c$
(Pounds  \et\ 2016), none being consistent with the wind velocity observed in 2001.  In a review of continuum-driven or
'Eddington' winds, King and Pounds (2015) furthermore showed that the observability of an individual, short-lived wind ejection may be of order months or less, as an
expanding shell presents a lower absorbing column to the nuclear x-ray source.

Considering the simultaneous observation of multiple expanding shells of absorbing gas in the context of super-Eddington accretion,  Pounds,
Lobban  and Nixon (2017) noted that short-lived wind components might be a natural  consequence of the way in which matter accretes in an AGN,
typically falling from far outside the radius of gravitational influence of the SMBH and with essentially random orientation. The resulting 
accretion stream will in general orbit in a plane misaligned to the spin of the central black hole (King \& Pringle 2006, 2007), with the inner
disc subject to Lense-Thirring precession around the spin vector, and orbits at smaller radii precessing sufficiently fast to cause
the tearing-away of independent rings of gas.

Computer simulations by Nixon \et\ (2012) then showed that - as each torn-off ring precesses on its own timescale - two neighbouring rings will eventually collide, with the
shocked material losing rotational support and falling inwards to a new radius defined by its residual angular momentum. A substantial increase in the local accretion rate might
then allow a previously sub-Eddington flow to become briefly super-Eddington, with excess matter ejected as a wind with velocity at or above the local escape velocity.
For typical AGN disc parameters, Nixon \et\ (2012) show the expected `tearing radius' is of
order a few hundred gravitational radii from the black hole - with predicted (escape) wind speeds in the range observed (King and Pounds 2015;
fig 4a).

The detection of multiple red-shifted X-ray absorption spectra during part of the same 2014 \xmm\ observation of \pg, corresponding to 
a substantial {\it inflow} of matter approaching the black hole at velocities up to
$\sim$ 0.3c (Pounds \et\ 2018: hereafter P18), offered the first observational support for the above accretion scenario. Simultaneous hard and soft X-ray spectra provided
independent confirmation of the high inflow velocity, with the soft X-ray absorption revealing
less ionized (higher density) matter, perhaps embedded in the main highly ionized flow. The authors note that disc tearing may be the origin of the high velocity, with inter-ring
collisions removing sufficient angular momentum for a near-direct infall to occur.
However, in that interpretation the shocked
gas is likely to be far too hot to be compatible with the rich absorption spectrum reported in P18, requring it to cool rapidly before being observed. Here,
we report evidence for that cooling radiation, identifying a variable X-ray continuum
component with the appropriate spectral characteristics and luminosity.
 
We assume a redshift for \pg\ of $z=0.0809$ (Marziani \et\ 1996), with  a black hole mass of $4\times 10^{7}$\Msun\ (Kaspi \et\
2000) indicating the historical bolometric luminosity 
is close to Eddington. 

\section{A broad continuum component dominating spectral variability}

While discrete spectral features provide a powerful diagnostic of the physical state and dynamics of matter in the inner accretion disc and surrounding AGN environment,
the continuum contains most of the X-ray luminosity. Conventionally, the AGN X-ray continuum emission is modelled by a power law, with typical photon
number index $\Gamma \sim$1.7, and a 'soft excess' parameterised by a black body with kT $\sim$0.1 keV. Both components are assumed to derive from disc accretion,
with the latter as thermal radiation from the hot inner disc and the harder power law resulting from up-scattering of disc photons in a hot electron corona
energised by magnetic field reconnection (analogous to the solar corona). Above $\sim$10 keV an additional broad X-ray 'reflection' component is often seen (Nandra and Pounds 1994),
as predicted from back-scattering of the hard power law flux by the accretion disc (eg George and Fabian 1991). 

In previous analyses of \xmm\ spectra of \pg, where discrete spectral features were the primary interest, a double power law provided an acceptable description of
the underlying X-ray continuum over the observable $\sim$0.4--10 keV spectrum (eg Pounds \et\ 2016), with continuum variability being dominated by the softer $\Gamma \sim$3
component. In the context of the present paper, it is interesting to note that a similar parametric description of continuum variability has been reported from extended
observations of other highly variable AGN, including the luminous Seyfert galaxies 1H 0419-577 (Pounds \et\ 2004) and MCG-6-30-15 (Vaughan and Fabian 2004).
In the former paper it was speculated that the variable continuum emission  might arise from shocks in outflowing gas, noting an alternative to the
variable power law in terms of thermal Comptonisation. More recently, a Principal Component Analysis of 26 \xmm\ spectra for a wide range of X-ray-bright AGN
(Parker \et 2014) found a variable-flux power law to be the common and dominant principal component of spectral variability.

To examine the nature of broad-band spectral variability in \pg\ over the 5-week \xmm\ observation of 2014, we first analysed 'difference spectra' between pairs of
individual orbits, confirming continuum variability to be dominated by changes in the normalisation of the soft ($\Gamma\sim$3) power law. To seek a more 'physical'
description of that variable continuum component, four flux-sliced 0.4--10keV spectra were created from the EPIC pn camera (Strueder \et\ 2001), across the full 2014 pn observation,
and chosen to contain a similar high
number ($\sim$ 400k) of counts in each flux slice. The resulting counts spectra for the high- and low-flux quartiles are shown in Figure 1, along with the subtracted
'difference spectrum'. 

\begin{figure}                                  
\centering                                                              
\includegraphics[width=6cm, angle=270]{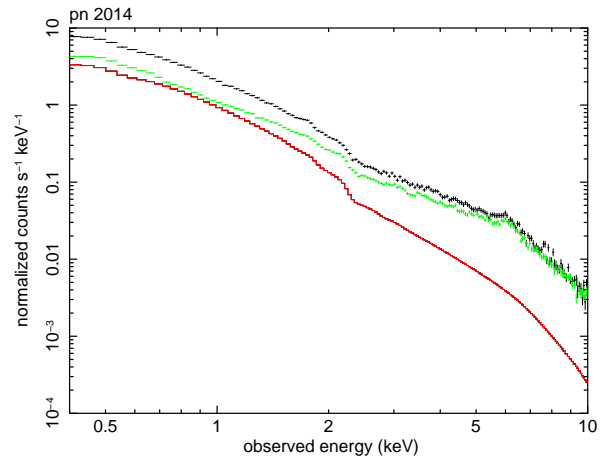}
\caption{High- and low-flux counts spectra obtained by flux-level slicing of the pn observation over the 2014 \xmm\ observation of \pg.
In red is a plot of the 'difference spectrum' obtained by subtraction of the mean low-flux spectrum from the mean high-flux
spectrum} 
\end{figure}

That difference spectrum was initially fitted over the restricted 1--10 keV band to minimise potential effects of variable soft X-ray absorption, using the xspec software package
(Arnaud 1996) version 12.10, the xspec model being TBabs*compTT, where TBabs fixed at  $3\times10^{20}$ cm$^{-2}$ represents Galactic absorption (Murphy \et\ 1986).
Free parameters in compTT
(Titarchuk 1994), a publically available model describing Comptonisation of soft photons in a hot plasma, are: the temperature of the soft photon spectrum, the (degenerate)
temperature and optical depth of the scattering electrons, and the normalisation of the output spectrum.

Thermal Comptonisation provided an excellent fit ($\chi^{2}$ of 140 for 142 d.o.f), with an input photon
temperature (kT=0.13$\pm$0.06 keV) consistent with thermal emission from the inner accretion disc. The degenerate pairing of electron temperature and optical depth for the scattering
matter is shown in Figure 2, where the confidence contours suggest a high initial plasma
temperature in the range $\sim$20--80keV, and corresponding electron scattering optical depth from $\tau_{es}$ $\sim$0.5--0.05.
In Section 3 we note that the much lower temperature ($\sim$ 1--2 keV) of the red-shifted absorber (P18) implies strong cooling if the observed infalling matter is
indeed from such a high temperature shock. 
  
\begin{figure}                                  
  \begin{center}
\includegraphics[width=6cm, angle=270]{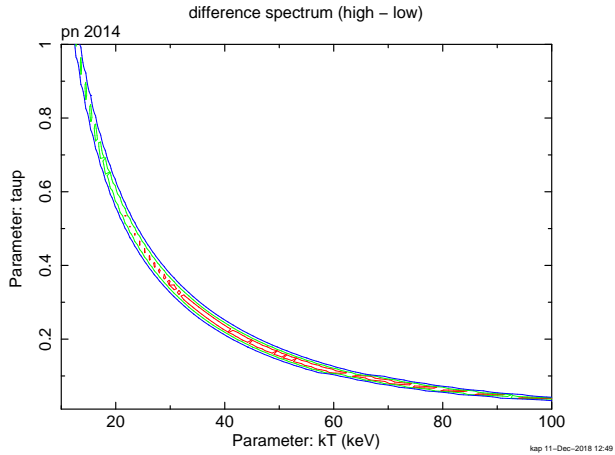}
\caption{Temperature versus optical depth for a Comptonised fit to the high-low 'difference spectrum' obtained by subtraction of the mean low-flux spectrum of PG1211+143.
The 95 percent confidence contour(red) indicates an initial plasma
temperature in the range $\sim$20--80keV, with corresponding electron scattering optical depth from $\tau_{es}$ $\sim$0.5--0.05.}
  \end{center}
\end{figure}

Extending the difference spectrum fit over the wider 0.4--10 keV band revealed a 'soft excess' and unacceptable overall  fit ($\chi^{2}$ of 186
for 156 d.o.f.). 
Adding a second Comptonised component, with the input photon temperature and scattering electron temperature tied across both components
recovered the excellent fit ($\chi^{2}$ of 152
for 154 d.o.f.), for a common
electron temperature of $\sim$43 keV and electron scattering optical depths of $\tau\sim$0.29 and $\tau\sim$0.12, respectively for the 'hard' and 'soft' spectral components, perhaps
indicative of spatial structure in the high temperature shock.
Over the observed (0.4--10 keV) band the luminosity of the hard and soft Comptonised emission components were $\sim$ 4.2$\times 10^{43}$ erg s$^{-1}$ and
$\sim$ 1.6$\times 10^{43}$
erg s$^{-1}$, respectively, with the latter component dominating the continuum emission below $\sim$1 keV.

In the context of shock cooling, the difference spectrum analysis suggests:
(1) a high initial shock temperature, with a lower limit an order of magnitude greater than the ionization temperature of the observed 0.3c infall;  
(2) an inhomogeneous cooling flow with effective scattering columns of $\tau_{es}$ $\sim$0.29 and $\tau_{es}$ $\sim$0.12; 
(3) a Comptonised spectrum consistent with exposure of the shocked gas to cooler photons from the accretion disc.

\section{Replacing the soft power law with thermal Comptonisation in the integrated highflux spectrum}
We found above that extending the Comptonisation fit to the highflux difference spectrum required a second component, suggesting a structured shock. Alternatively, as we see below,
inclusion of absorption and emission
features as seen in the stacked 2014 spectrum of \pg\, provides a good fit over the whole
0.4-10.0 keV spectral band with a single Comptonised component.

We exploit the high
statistical quality of the integrated highflux data, with a total exposure of 98ks and 583000 counts over the 0.4-10 keV spectral band, to explore the use of alternative (power law and thermal
Comptonisation) continuum components in a full spectral model.
Our starting point is the spectral
model reported in Pounds \et\ (2016) from an analysis of the stacked pn data from the \xmm\ 2014 observation. 

\begin{figure}                                  
\centering                                                              
\includegraphics[width=6cm, angle=270]{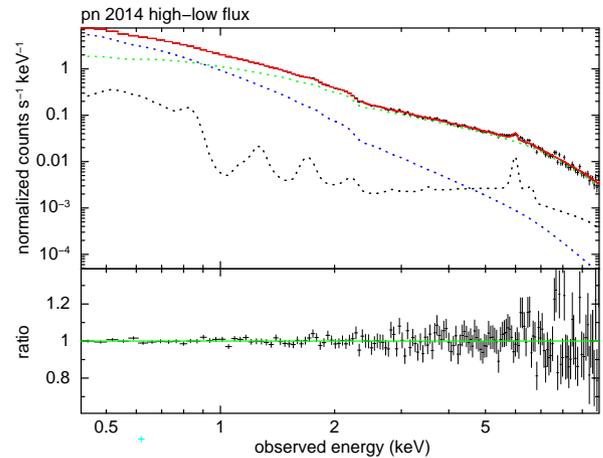}
\caption{Continuum fit to the highflux pn data} 
\end{figure}

Those earlier analyses focussed on the detection of absorption line spectra (eg. Pounds \et\ 2016), revealing high velocity
ionized outflows in line-of sight to the hard x-ray source. The underlying X-ray continuum was adequately modelled by a hard power law ($\Gamma\sim1.65$) and black body (kT$\sim$0.1
keV), together with a softer power law ($\Gamma\sim3$) responsible for most of the inter-orbit variability. Here we use the same spectral model to fit the stacked high-flux data from 2014, but
replacing the soft power law with compTT, to see if that more physical description can be distinguished when the variable continuum component is strongest.

The complete spectral model in xspec is then: TBabs(po+compTT+xillver+ph-em)*ph-abs, where xillver (Garcia \et\ 2013) models the continuum reflection (hard continuum and fluorescence line emission),
and ph-em and ph-abs represent the absorption and re-emission from the photoionized wind. It is interesting that no
additional black-body component is now required. Fitting the high-flux data over the full
spectral band 0.4--10 keV yielded an excellent fit ($\chi^{2}$ = 136 for 149 d.o.f.). Figure 3 shows the model spectrum, together with the data-model residuals. Over the more
restricted 1--10 keV band the model parameters were very similar with a further good fit ($\chi^{2}$ = 132 for 132 d.o.f.)

\begin{figure}                                  
\centering                                                              
\includegraphics[width=6cm, angle=270]{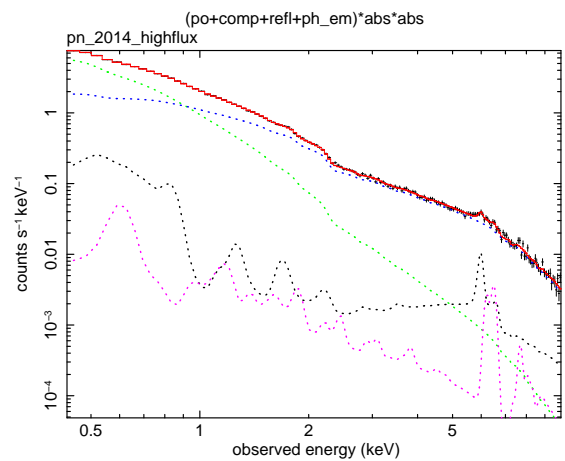}
\caption{(top)Full spectral fit to the 2014 pn highflux data, with parameters listed in Table 1. Colour coding is: total emission (red), hard power law (blue), compTT (green), ionized
reflection (black), photoionized emission (magenta)} 
\end{figure}

Replacing the compTT continuum component with a second power law, as used in the earlier analyses, yielded a near identical fit over the 1--10 keV band ($\chi^{2}$ = 132 for 134
d.o.f.), but extending the model to 0.4 keV was significantly worse ($\chi^{2}$ = 164 for 151 d.o.f.), an f-test finding the Comptonised fit preferred with 99.9$\%$ confidence, with
the data-model residuals indicating the intrinsic curvature of the compTT component is responsible for the significantly better broad band fit - while removing the need for significant soft
x-ray emission
from the inner accretion disc(and the related black body component in a spectral model).

\section{Discussion}

The detection of ultra-fast highly ionized outflows (UFOs) in early x-ray spectra of the narrow line Seyfert galaxy \pg\ (Pounds \et\ 2003) and
the luminous QSO PDS 456 (Reeves \et\ 2003), opened up a new field of study of AGN, well suited to the uniquely high throughput of x-ray
spectrometers on ESA's \xmm, launched in late 1999. King and Pounds (2003, 2015) noted that such winds are a natural result of a high accretion ratio,
with excess matter being driven off by radiation pressure when the accretion rate exceeds the local Eddington limit (Shakura and Sunyaev
1973; hereafter SS73).

While this picture provided a satisfactory explanation of most UFOs, where a single detection yielded a unique wind velocity, the extended
study of \pg\ in 2014 found a more complex outflow profile, with velocities of $\sim$0.06c, $\sim$0.13c and $\sim$0.18c detected in the
stacked data set. Such complexity was clearly inconsistent with a wind profile launched from a flat axi-symmetric disc (SS73), suggesting some intrinsic disc
instability or rapidly variable accretion rate  being introduced to the inner disc.

The detection of multiple red-shifted X-ray absorption spectra during part of the same 2014 \xmm\ observation of \pg, where the extreme redshift ($\sim$0.48) indicated absorption in  
a substantial {\it inflow} of matter approaching the black hole at a velocity of
$\sim$ 0.3c (Pounds \et\ 2018: hereafter P18), offered the first observational support for such rapidly rapidly variable accretion.
While the inflow observation was of high significance in only one of seven individuel observations, inflows of lower column density and smaller redshift were detected in 5 of the other 6
EPIC pn observations, with
redshifts ranging from 0.20 to 0.36 (and line-of-sight inflow velocities from 0.1 -- 0.23c).    

As noted in the Introduction, periods of highly variable accretion to the inner disc could  be a
consequence of the way in which AGN accrete, where gas initially falls towards the SMBH with essentially random orientation. 
Lense-Thirring precession would cause misaligned orbits to precess around the black hole spin vector, with the innermost ring(s) warping and potentially breaking
off.  Collision between 
neighbouring rings, rotating at different rates, would then shock, with loss of rotational support leading to matter falling freely to a new radius defined by
its residual  angular
momentum, where - if it can cool fast enough - it forms a new disc (Nixon \et\ 2012).

In this paper we have presented an analysis of the energy-dominant x-ray continuum for \pg, finding that strong inter-orbit variability over the 7-weeks \xmm\ study in 2014
was dominated by variability of a soft ($\Gamma \sim 3$) power law component. Further comparison of flux-sliced spectra confirmed the normalisation of the soft power law to be the
dominant variable.

In seeking a more physical description of that variable continuum component, a thermal Comptonisation model was used to replace the soft power law, finding an excellent spectral fit to the
high-flux data slice, over the full 0.4--10 keV observed spectrum, when photoionized emission and absorption spectra were included.

An upper limit to the initial shock temperature, corresponding to opposing rings colliding head on, with the kinetic energy of opposing Keplerian flows being fully thermalised, would be
of order T$_{shock}$ $\sim GMm_{p}$/kR,
for colliding rings of radius R (R$_{g}$), and a black hole mass M(M$_{sun}$).
For \pg\ and R$\sim$500, T$_{shock}$ $sim 10^{10}$ K, which is uncomfortably high in comparison with the temperature range found from thermal Comptonisation fit to the
variable X-ray continuum component. However, the assumption that the kinetic energy is fully thermalised seems unlikely, since the initial
shocked plasma will be highly compressed and optically thick, with rapid adiabatic expansion lowering the temperature and density before being subject to the further (Compton)
cooling by soft
photons from the accretion disc. Here we consider the variable x-ray continuum component as evidence for that further cooling, providing a quantitative link
between the Compton-cooled matter and a line-of-sight element of the subsequent inflow observed in x-ray absorption (P18).

The detection of strong K-shell absorption in Si and S (as well as Fe, Ca and Ar) indicates an inflow ionisation temperature of 1-3$\times10^{7}$K
(Shull and Steenberg 1982), with this being an upper limit to the electron temperature since - as assumed above - photoionisation is likely to be dominant for matter exposed to the inner
accretion disc. However, the Comptonisation spectral fits reported here indicate an initial temperature {\it at least} a factor 10 higher.

P18 estimate an accretion rate for the 0.3c infall of $\sim 10^{23}$ gm s$^{-1}$, yielding a luminosity of $\sim 10^{43}$ erg s$^{-1}$
for a mass/energy conversion efficiency $\eta\sim$0.1, and lasting for $\sim 10^{5}$ s. 
In comparison, the mean bolometric luminosity of \pg\ is $\sim 10^{45}$ erg s$^{-1}$, with $\sim 10^{44}$ erg s$^{-1}$ arising from the inner disc region relevant here.
Variability of the Comptonised spectral component on a timescale of hours (Fig. 5), suggests a small number of
infall events - similar to that observed - can make a significant contribution to the inner disc accretion rate. But how does that compare quantitatively with the putative
cooling radiation?

\begin{figure}                                  
\centering                                                              
\includegraphics[width=8cm]{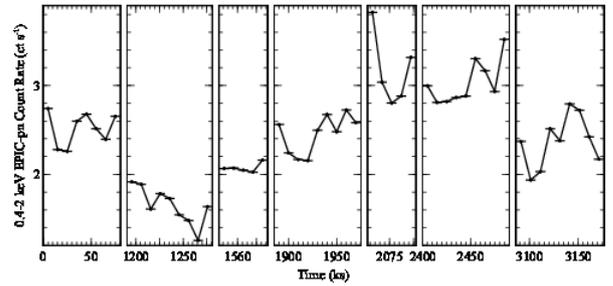}
\caption{EPIC pn count rate at 0.4--2 keV, the spectral band dominated by the soft Comptonised continuum, summed in 10 ks bins.} 
\end{figure}

Modelling of the 0.3c inflow in P18 found an ionization parameter $\xi$ $\sim$3000 erg.cm.s$^{-1}$ and absorbing gas column density N$_{H}$ $\sim 4.2\times10^{23}$
cm $^{-2}$. The assumed free-fall of the absorbing matter provided an estimate of its radial distance (r) from the source of hard x-ray ionizing radiation
(L$_{ion}$), allowing the particle density (n) to be estimated, since - for a photoionized gas - $\xi$ = L$_{ion}$/n.r$^{2}$). The ionizing luminosity relevant to
the crucial FeK spectrum, determined from the continuum fit (3$\times 10^{43}$  erg s$^{-1}$), then yields a particle density n$\sim$ $2.5\times10^{11}$ cm$^{-3}$.

We assume the luminosity of 'cooling radiation', averaged over 5 weeks,
provides a measure of the mean rate of shocked matter available to accrete, both in and out-of the line of sight.  
We further note that for a shock collision at radius R (R$_{g}$), with y percent of the thermal energy subsequently emerging as soft x-radiation, and the subsequently cooled matter then
accreting to the inner disc, the ratio of radiation efficiencies would be of order 100R/6y for zero black hole spin.

Comparing the observed cooling luminosity of $\sim$ 6$\times 10^{43}$ erg s$^{-1}$ to L$_{bol}$ requires 
shocks at 500 R$_{g}$ (as assumed) for the resulting matter accretion rate to be compatible with the inner disc luminosity. While the presence of co-moving, more highly ionized (transparent)
matter in the inflow, helpful in maintaining the dynamical stability of the accreting stream, together with less efficient shocks, could affect this comparison, the putative link seems very
plausible.

Finally, we note that if disc tearing is indeed an important source of direct inner-disc accretion in luminous AGN, then both ultra-fast inflows and Comptonised cooling radiation should
also be common. Is there such evidence?

As mentioned in the Introduction, similar difference spectra to that found in \pg\ have been reported previously for two other highly (x-ray) variable, luminous Seyfert galaxies,
1H0419-577 (Pounds \et\ 2004) and MCG-6-30-15 (Vaughan and Fabian 2004). More recently, a principal component
analysis of spectral variability in 26 AGN suggested a soft 'power law' continuum may be a common feature (Parker \et\ 2014).
Linking the Comptonised emission reported in the present paper with the recent evidence for fast inflows potentially offers a physical explanation of that principal component.

In summary, we find the principal component of X-ray variability in the luminous Seyfert galaxy \pg\ - interpreted here as thermal Comptonisation of off-plane shocked gas - has x-ray 
spectral characteristics, variability and luminosity consistent with an origin where disc tearing is a major contributor to the accretion process in AGN.

\section{Data availability}

The data underlying this article are available in the XMM archive at http://nxsa.esac.esa.int/nxsa-web.

\section*{ Acknowledgements }
\xmm\ is a space science mission developed and operated by the European Space Agency. We acknowledge in particular the excellent
work of ESA staff in Madrid successfully planning and conducting the \xmm\ observations.

\bsp	
\label{lastpage}

\end{document}